\documentstyle[11pt,newpasp,twoside,epsf]{article}
\begin{document}

\title{The metallicity of star-forming gas over cosmic time}
\author{Simon J. Lilly \& C. Marcella Carollo}
\affil{Department of Physics, Swiss Federal Istitute of Technology (ETH-Z\"urich), ETH-H\"onggerberg, CH-8093 Z\"urich, Switzerland
}
\author{Alan N. Stockton}
\affil{Institute for Astronomy, University of Hawaii, 2680 Woodlawn Drive, Honolulu, Hawaii 96822, USA}

\begin{abstract}
A brief review concludes that there is now good overall 
agreement between theoretical estimates of the energy associated with the
production of the observed metal content of the Universe and
the observed extragalactic background light. In addition the overall
form of the star-formation history over $0 < z < 5$ is reasonably
well constrained.  

The study
of emission line gas in galaxies as a function of redshift provides a complementary view of
chemical evolution to that obtained from studies of absorption line systems in quasar spectra.
Emission line gas is more relevant for some questions, including the
confrontation with models for the chemical evolution of our own and other galaxies,
and Origins-related questions about the formation of stars and planets.
Using relatively crude diagnostic parameters such as Pagel's $R_{23}$, 
the increase of the metallicity of star-forming gas with cosmic 
epoch can be tracked.  Observations of a large sample of CFRS galaxies at $z \sim 0.8$
(a look-back time of 0.5$\tau_0$)
show the appearance of a significant 
number of low metallicity systems
amongst luminous L* galaxies. However, overall there is only a 
modest change in mean
metallicity compared with the present, $\Delta$ log $Z = -0.08 \pm 0.06$. 
These data do not support a 
fading-dwarf scenario for the ``faint blue galaxies''. 
At $z \sim 3$ it is likely all star-forming 
galaxies have sub-solar metallicities.  The overall increase in 
the metallicity of star-forming gas with cosmic epoch matches rather
well the age-metallicity relation in the solar
neighbourhood and some recent models for the evolution of the global metallicity 
of gas in the Universe.

\end{abstract}

\section{Introduction}

As is clear from the theme of this conference, the production of the 
heavy elements through
stellar nucleosynthesis is central to the Astronomical Search 
for Origins.    Clearly, the metallicity of the Universe and of objects in it
provides a 
fundamental metric reflecting the development of structure and
complexity in the Universe on galactic scales. This metric is 
all the more 
important because it is
relatively easily observable and ``long-lived'' in the sense
that heavy atomic nuclei, once produced, are not readily destroyed.  Thus, the
metallicity of material reflects its entire evolutionary
path through cosmic time - with a potential for sophistication which
is impressive (see Jim Truran's contribution to these proceedings).   
Furthermore, in regard to the Origins theme,
it is these same heavy elements that play a key role in the formation of
structure on planetary scales and which are fundamental to the
existence of Life itself in the Universe.

For many years, the metallicities of gas at high redshifts have been 
studied through the analysis of absorption line systems seen in quasar 
spectra.  These are discussed by Wal Sargent in these proceedings.
The lines of sight to quasars probe, almost by definition, 
random regions of the Universe - the only possible concern being whether lines
of sight passing through dusty regions are under-represented
because the
quasar is then eliminated from the sample (Fall and Pei 1993, see Ellison et al 2001).  
On the other hand, only a single
line of sight through a given system is generally 
available, making the interpretation of the metallicity measurement in the context
of larger structures, such as galaxies, non-trivial and fundamentally
statistical in nature.  As an example, the
relationship between the high column density ``damped Lyman $\alpha$'' (DLA) 
systems and galaxies is still by no means clear.

The study of the metallicities of material in known galaxies
at high redshift is at a much earlier stage of development and is inevitably
of lower sophistication.
However, metallicity 
estimates, especially of the [O/H] abundance, using diagnostics that are
based solely on strong emission lines,
e.g. [OII] 3727, H$\beta$, [OIII] 4959,5007, H$\alpha$, [NII]6584, [SII]6717,6731,
are now technically feasible over a wide range of redshift and give a new perspective 
on the chemical evolution
of galaxies.  

For some purposes, the estimates of the metallicity of star-forming gas
at earlier epochs that are obtained from emission lines may actually 
be of more relevance than the more
``global view'' that is obtained from the absorption line studies. 
Applications include the confrontation of
models for the chemical evolution of the Milky Way or other
galaxies, the comparison with the observed age-metallicity relationship seen
in galaxies at the present epoch, and the use of metallicity estimates to constrain
the present-day descendents of high redshift galaxies.
The emission line gas in star-forming regions is also the
most relevant for the planetary and astro-biological aspects of the
Origins theme since it should be 
representative of the material out of which the stars and planets are 
actually being made.

In this contribution, we first review
the production
of metals in the Universe in the context of its stellar content and the
extragalactic background light (EBL).  We then
describe and discuss measurements of star-forming gas 
at different cosmic epochs, focussing on new results that we have obtained from a large sample
at $z \sim 1$. 
Throughout the discussion we adopt the so-called ``concordance'' cosmology with
H$_0 = 70$ kms$^{-1}$Mpc$^{-1}$, $\Omega_0 = 0.3$ and $\lambda_0 = 0.7$.

\section{Global considerations: the EBL and the production of metals}

\subsection{The inventory of metals}

Several authors have constructed estimates for the global metal content of the
Universe (e.g. Fukugita, Hogan \& Peebles 1998, Pagel 2002). The recent determination 
of the galaxy luminosity function from the Sloan Digital Sky Survey (Blanton et al 2001)
has increased the local luminosity density by 50\%. Many of the components
will scale with this quantity. 
The entries 
in the following table are 
derived from those of Fukugita et al (1998) and Pagel (2002) with an SDSS-inspired increase
applied to all but the hot cluster gas.

\begin{center}
\begin{table}
\caption{Inventory of nucleosynthetic products in the Universe}
\begin{tabular}{llll}
\tableline
Component & baryons($\rho/\rho_c$ & $Z/Z_{\sun}$) &  metals($\rho/\rho_c$) \\
\tableline
\multicolumn{4}{l}{{\bf Accesible heavy element components}} \\
\tableline
Spheroid stellar atmospheres & $0.004 h_{70}^{-1}$      &1.0& $7 \times 10^{-5} h_{70}^{-1}$ \\
Disk stellar atmospheres    & $0.0015 h_{70}^{-1}$      &1.0& $3 \times 10^{-5} h_{70}^{-1}$ \\
Hot cluster gas       & $0.0025 h_{70}^{-1.5}$    &0.3  & $1.6 \times 10^{-5} h_{70}^{-1.5}$ \\
Warm group/field gas  & $0.02 h_{70}^{-1.5}$?    &$\le$ 0.3?  &$\le 1.2 \times 10^{-4} h_{70}^{-1.5}$? \\
\tableline
Sum                   & $\sim 0.028 h_{70}^{-1}$   &  & $1.2-2.4 \times 10^{-4} h_{70}^{-1}$ \\
\tableline
\multicolumn{4}{l}{{\bf Inaccessible components}} \\
\tableline
$^4$He in stellar cores         &                           & all $^4$He & $\sim 0.8 \times 10^{-4} h_{70}^{-1}$ \\
\tableline
\tableline
\end{tabular}
\end{table}
\end{center}

The final entry in the Table reflects the $^4$He that has been produced
in stellar cores below 10 M$_{\sun}$ and which remains locked up in them (including white dwarfs).
Assuming that all stars process about 10\% of their starting material 
into $^4$He, the $^4$He that has been produced after 10 Gyr should be of order 
0.04 of the total remaining
stellar mass (for a Salpeter-like mass function).

Several well-known points are evident in this table (the reader is referred to
the discussions in 
Fukugita et al 1998, Pagel 2002 and also Bernstein, Freedman \& Madore 2002).    
First, although some of the most important components are quite uncertain,
the ``best-guess'' sum of the accounted baryons is now close to that predicted by
Big Bang Nucleosynthesis ($0.03 h_{70}^{-2}$).
The largest single component
of the baryon budget is the warm gas outside of galaxies in small groups and is
unfortunately its most uncertain, both in terms of baryons and metals.
Second, the stars in spheroids outnumber those in disks by a significant factor. 
This should alert us to the
possibility that
modes of star-formation not seen in our own Galaxy today may have dominated star-formation
(and light production)
in the Universe at earlier epochs.
Finally, the addition of substantial metals outside of galaxies without additional stars 
means that the effective yield has been higher (0.025-0.05) than has been estimated from the
Solar neighbourhood.

\subsection{The production of light}

Gravitational collapse and stellar nucleosynthesis are the only two
processes thought to generate significant amounts of luminous energy in the Universe
(excluding exotic possibilties such as decaying particles etc.).
The relationship between the production of light and metals is set by fundamental physics:

\begin{equation}
\rho_{L,bol}(z) = 0.007 (1+\alpha) \beta^{-1} \Delta\rho_Z(z) c^2
\end{equation}

In eqn. (1), 0.007 is the nucleosynthetic efficiency, the $(1+\alpha) \sim 2.6$ is a term to 
account for the $^4$He production
associated with the production of the heavier elements, $\rho_Z$, and the
$\beta$ term (of order unity)
is to account for metals
``lost'' in neutron stars and blackholes.
Once produced at a redshift $z$, the
appearance today of a given component of the radiation field is also very simple. In
bolometric terms:

\begin{equation}
I_{EBL,bol} = {{\rho_{L,bol}(z) c}\over{4\pi(1+z)}} 
\end{equation}

The form of eqn.(2) has some interesting consequences.  For a ``flat-spectrum'' background with
constant $I_\nu$ and some frequency cut-off, $\nu_{max}$, as would be approximately produced by blue star-forming
galaxies longward of the Lyman limit, the monochromatic 
surface brightness of the background is
independent of the redshift at which it is produced, for as long as 
$\nu_{obs} \le (1+z)\nu_{max}$,
since the $(1+z)$ term is cancelled by the band-width term in the $k$-correction.
Thus there is a one-to-one correspondence between metal production and
the contibution to such a
flat-spectrum background (Lilly \& Cowie 1987).
It is then easy to see
that if the background
is produced with $I_{\nu} \propto \nu^{\alpha}$, then 
the redshift distribution of photons in the background in this spectral region
will be weighted, relative
to the redshift history of star and metal production, by a potentially large factor of $(1+z)^{\alpha}$. 
As an example, SCUBA submm-selected samples at 850 $\mu$m (which have $\alpha \sim 3.5$) are heavily 
biassed to
high redshifts relative to the cosmic star-formation history (Lilly et al 1999) and thus
mostly have $z >> 1$. 
On the other hand, the ISO-selected samples at 15$\mu$m are heavily biassed against high redshift
objects and mostly have $z < 1$ (e.g. Chary and Elbaz 2001).

The $(1+z)^{-1}$-weighted average redshift for the production of the background 
is probably $z \sim 1$. As an example,
a star-formation history that rises as $(1+z)^3$ to $z = 1$ and is constant at higher redshifts 
gives an average
$z = 1.14$. Assuming $z \sim 1$,
it is then
easy to compute the background that should be associated with each component in Table 1, noting that
for the hidden $^4$He in stellar cores, the (1+$\alpha$) term is not applied.  These are shown in Table 2.

In addition, the likely component due to accretion on to blackholes in 
AGN may be simply calculated from the estimated density of black-holes
(see e.g. Fabian \& Iwasawa 1999).

\begin{equation}
I_{EBL,bol} = {{\rho_{BH} \eta c^3}\over{4\pi(1+z)}}
\end{equation}

Here $\eta$ is the radiative efficiency of black-hole accretion ($\eta \sim 0.1$).
The present day global density of blackholes is about 0.5\% of the density of spheroids
(Magorrian et al 1998).
With $z \sim 2$ (for quasars) and the pre-SDSS density of the spheroids (which is likely 
most appropriate
for the Magorrian calibration)
this gives an additional non-nucleosynthetic component
to the EBL of about $10 h_{70}^{-2}$ nWm$^{-2}$sr$^{-1}$.

\begin{center}
\begin{table}
\caption{Predicted production of the EBL from elements of Table 1}
\begin{tabular}{lc}
\tableline
Component & Predicted EBL intensity (nWm$^{-2}$sr$^{-1}$)\\
\tableline
Stellar atmospheres       &  $27 h_{70}^{-1}$ \\
Hot cluster gas           &  $4 h_{70}^{-1.5}$  \\
Warm group/field gas      &  $32 h_{70}^{-1.5}$ \\
$^4$He in stellar cores   &  $23 h_{70}^{-1}$ \\
AGN                       &  $10 h_{70}^{-2}$ \\
\tableline
Sum                       &  $\sim 96 $   \\
\tableline
\end{tabular}
\end{table}
\end{center}

It should be noted that the predicted total
estimate in Table 2 of around 100 nWm$^{-2}$sr$^{-1}$ is higher than, for instance, that in Pagel (2001) because
of (a) the higher SDSS luminosity density folding through the overall stellar density, this 
representing a 50\% increase; and (b) the choice of
$z \sim 1$ for the $(1+z)^{-1}$-weighted average redshift of production of the background,
another 50\% increase over the $z \sim 2$ adopted by Pagel.

\subsection{The extragalactic background light}

The bolometric content of the EBL is observationally
quite well constrained.  
Aside from the primordial Cosmic Microwave Background, the EBL is dominated by two components peaking around 1 
$\mu$m and around 100-200 $\mu$m respectively (Fig 1).  Longward of 6 $\mu$m, the total energy content is well-defined
from COBE to be
40-50 nWm$^{-2}$sr$^{-1}$ (e.g. Hauser \& Dwek 2001).  

In the optical, a firm lower bound at 20 nWm$^{-2}$sr$^{-1}$
is obtained by integrating the observed galaxy number counts. 
Recently Bernstein et al (2002) have claimed a significant detection of an integrated 
background in the optical that is substantially higher by a factor of 2-3.  Much of the discrepancy
is likely associated with low surface brightness parts of detected galaxies whose central
high surface brightness regions have been already detected.  Using a 
reasonable model for the missing parts of the spectrum, they estimate a total bolometric EBL 
over the whole wavelength range out to 1 mm of $100 \pm 20$ nWm$^{-2}$sr$^{-1}$.

The agreement of the observed EBL with the total 96 nWm$^{-2}$sr$^{-1}$ predicted for $h_{70} = 1$
in Table 2 is quite satisfactory. 

\begin{figure}
\plotone{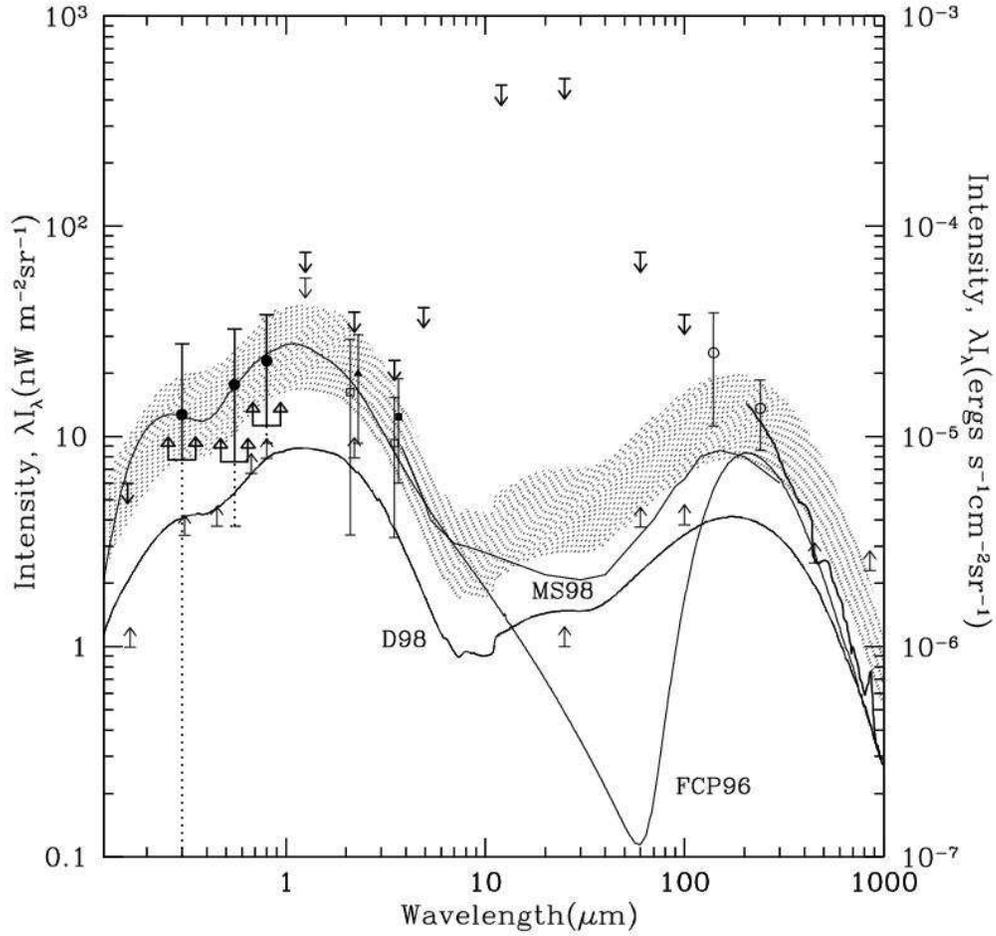}
\caption{Measurements and limits on the EBL, reproduced from Bernstein et al (2002)
which should be consulted for the references for individual data points.  
The shaded region is a scaled version of the model by Dwek et al (1998).  The increase
in the optical EBL claimed by Bernstein et al makes the energy content in the ``obscured''
far-IR and ``unobscured'' (ultraviolet, optical and near-IR) components roughly equal.
}
\end{figure}

\subsection{The redshift production of the background}

Resolution of the background into astrophysical objects of known redshifts, luminosities
and so on, would enable us to understand the whole history of metal production.   

A great deal of attention has recently been paid to estimates of the luminosity density
of the Universe at different wavelengths as a function of redshift.   In the optical (rest-frame ultraviolet), 
the initial rise to $z \sim 1$ seen in the CFRS (Lilly et al 1996) has remained reasonably robust
although its steepness, originally thought to be $(1+z)^4$, is reduced by approximately one
power of $(1+z)$ by the change in cosmology and would be further flattened by another half power 
if the SDSS results for the local luminosity density are correct.
At higher redshifts it now seems likely that the luminosity density is roughly constant with
redhift to $z \sim 5$ (see Sawicki et al 1997, Steidel et al 1999) rather than declining as originally proposed (Madau et al 1996).
Some light has likely been
missed at high redshifts on account of 
surface brightness selection (e.g. Lanzetta
et al 2002) and the discrepancy of the ``observed'' counts with the integrated EBL measurement
of Bernstein et al (2002) also suggests this.  However, it is quite likely that the
missing light is associated with the fainter detected objects 
and so the form
of the ultraviolet luminosity density with redshift may well be quite similar to that 
previously claimed
out to redshifts as high as $z \sim 5$.

In the far-IR, where the 
total background is better constrained at $\lambda > 1 \mu$m, much less of the 
background has hitherto been resolved
and the redshift distribution is less certain.
At 15 $\mu$m, deep counts are available from 
ISOCAM surveys and Chary and Elbaz (2001) show that the rapid evolution required to account
for these must flatten at $z \ge 1$ so as not to overproduce the background seen
beyond $100 \mu$m (see also Lagache, Dole \& Puget 2002).   This is broadly similar to the 
behaviour seen in the 
ultraviolet, with a possibly steeper rise at low redshifts.  Fig 2 reproduces a
figure from Chary \& Elbaz, which is probably reasonably close to reality.

\begin{figure}
\plotone{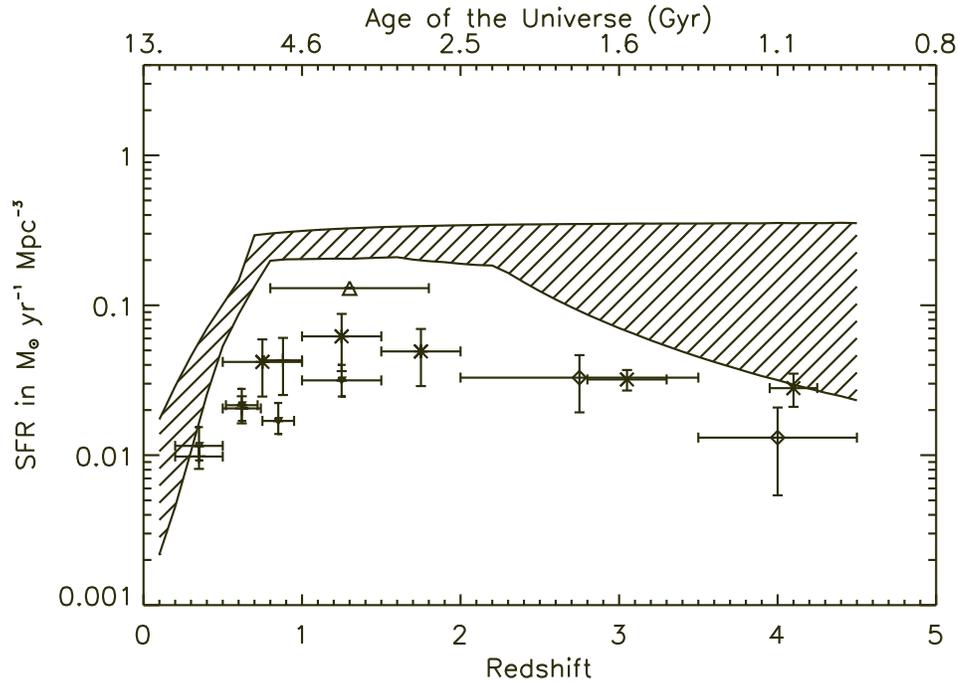}
\caption{The Universal star-formation rate, taken from Chary \& Elbaz (2001). The data points show estimates 
of the ``unobscured''
star-formation rate obtained from the ultraviolet luminosity density as f(z).  At high redshifts the estimates are probably
low by a factor of a few 
because of surface brightness selection effects as evidenced by the fact that the observed galaxies do not
account for the integrated EBL (Bernstein et al 2002), as well as more direct arguments (Lanzetta et al 2002).
The hatched area show the range of acceptable models (Chary \& Elbaz 2001) 
for the far-IR background that are consistent with 
existing ISOCAM and SCUBA surveys and with the shape of the far-IR background.  It should be noted that 
this figure has been
computed for an $\Omega = 1$ cosmology.}
\end{figure}

\section{Measurements of the metallicity of star-forming gas over cosmic time}

As noted above, the measurement of the metallicities of star-forming gas in galaxies at 
different epochs is complementary to the study of the metallicities of gas generally in the 
form of absorption line systems.  One
advantageous feature is that diagnostics have been developed using spectral features in the
visible $3700 \le \lambda \le 6600 \AA$ waveband.  These may be followed to high redshifts
in a uniform way
by working in the infrared waveband.

The $R_{23}$ parameter introduced by Pagel (1979) provides an estimator for the [O/H] abundance.
$R_{23}$ has some undoubted drawbacks - it is sensitive to reddening and has two metallicity solutions for
most values of $R_{23}$. It also depends on ionization (though this can be determined relatively 
easily). These issues have been amply discussed in the literature (see e.g. Kobulnicky,
Kennicutt \& Pizagno 1999,  Kewley \& Dopita 2002).  
It's great virtue is that it is based on a few strong emission lines and it can
therefore be applied in a uniform way over a wide range of redshifts.  Some of the difficulties 
in calibration can be circumvented
by considering relative effects rather than relying on absolute determinations of metallicity.

There have been many applications of $R_{23}$ in the local Universe, e.g. van Zee et al's (1998)
extensive study of HII regions in nearby galaxies spanning a wide range of metallicities.
At high redshift, we are restricted at present to measurements of the integrated light from the galaxies.
In this regard, the sample of $\sim 200$ nearby galaxies observed with slit-scanning techniques
by Jansen et al (2000, 
hereafter JFFC)
is especially useful and serves as a basic reference point for studies at higher redshifts.

\subsection{New measurements of CFRS galaxies at $0.5 < z < 1$} 

We have been undertaking a program of spectroscopy of star-forming galaxies selected from the 
Canada-France-Redshift Survey (CFRS; Lilly et al 1995)
at intermediate redshifts $0.5 < z < 1.0$.
The first results of this (based on a sample of 15) were reported in Carollo \& Lilly (2001).
We have now observed about 100 galaxies from which we can form a
statistically complete sample of 65 galaxies.  Our new sample is defined to have
$0.47 \le z \le 0.92$, $I_{AB} \le 22.5$ (which means that the galaxies are 
roughly $L*$ or greater in
luminosity, $M_{B,AB} \le -20$) and rest equivalent width of H$\beta \ge 8 \AA$. These galaxies 
are representative of 90\% of the [OII] 3727 luminosity in the CFRS at these redshifts, and
probably account for a half of all of the not-heavily-obscured star-formation at these
epochs (Lilly et al 1996).  Other relevant studies that have been published recently 
include those of Kobulnicky \& Zaritsky (1999:
14 galaxies $0.1 < z < 0.5$)
Hammer et al (2001: 14 CFRS galaxies $0.45 < z < 0.8$ and Contini et al (2002: 68 mostly local but ultraviolet
selected galaxies).  A more detailed discussion of our new sample
and the conclusions that we have drawn from it may be found in Lilly, Carollo \& Stockton (2002, in preparation).

In terms of the simple diagnostic ratios $R_{23} = (f_{3727} + f_{4939+5007}) / f_{4861}$
and $S_{32} = f_{4959+5007} / f_{3727}$, the galaxies in this sample are very similar 
to those in the local JFFC sample (see Fig 3). They are probably a little more reddened but
this is expected given their higher mean luminosities - the JFFC sample extends 
over $-14 \le M_B \le -23$.  Unfortunately we do not at present have H$\alpha$ spectrophotometry
enabling us to apply individual de-reddening, and so have applied a uniform $E(B-V) = 0.3 \pm 0.15$
derived from the JFFC sample at similar luminosities.   The dual value degeneracy 
of $Z$ with $R_{23}$ 
may be broken by using the [NII] 6584 line, in combination with either H$\alpha$ (for which the line ratio
is reddening independent)
or with one of the oxygen lines at shorter wavelengths (see e.g. Kewley \& Dopita 2002).  
All the 6 objects for which we have [NII]/H$\alpha$
have metallicities on the higher $Z$ branch for $R_{23}$ and we have assumed that this
is the case for all of the objects.  New observations of H$\alpha$ and [NII]6584 are now scheduled
on the VLT so these uncertainties should be reduced considerably in the future. 

\begin{figure}
\plotone{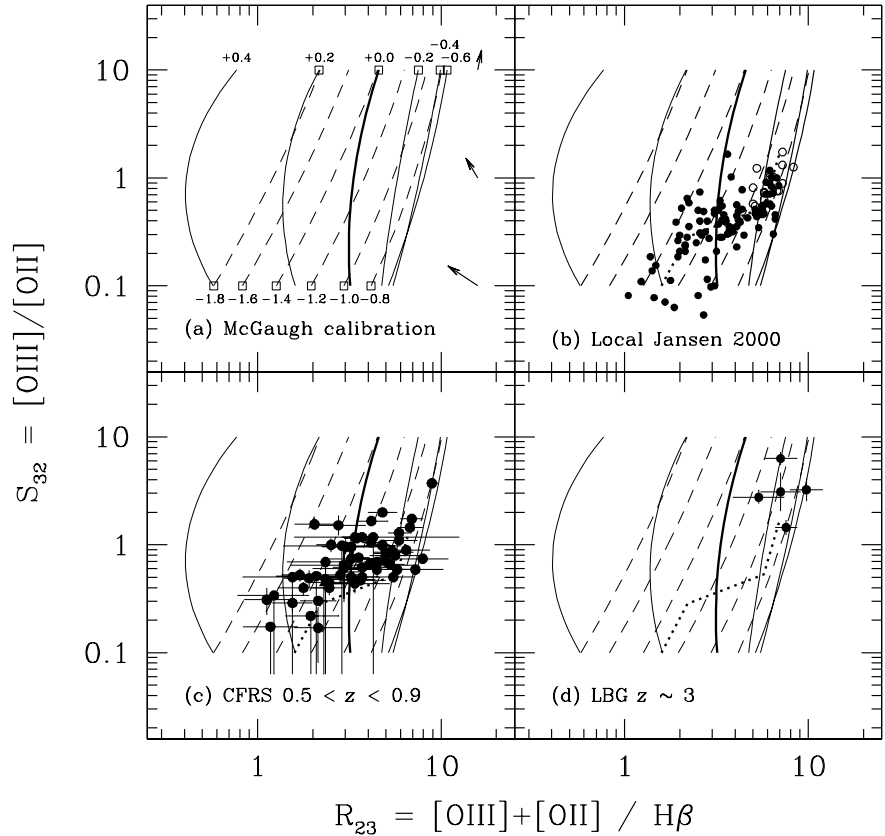}
\caption{$R_{23}$-$S_{23}$ diagnostic diagram.  Lines of constant metallicity are approximately
vertical and move first to the right (dashed lines) 
and then to the left (solid lines) as the oxygen abundance increases,
turning around at $Z \sim 0.25 Z_{\sun}$. The McGaugh (1991) calibration is shown in Panel (a).
Panels (b)-(d) show the JFFC ($-14 < M_{B,AB} < -23$), CFRS and LBG samples respectively. The dotted line
represents the ridge line in the JFFC sample.  None of the points are corrected for 
reddening, and it is likely that the small offset of the CFRS with respect to 
the JFFC is due to 
a higher average reddening in the CFRS.   Objects on the lower $Z$ branch in JFFC are
represented by open circles.  Star-forming galaxies at $z \sim 0.8$ appear very similar to 
their low redshift counterparts, while the LBG at $z \ge 3$ have higher ionizations.}
\end{figure}

Carollo \& Lilly (2001) emphasized the lack of change in $R_{23}$ and, by implication,
in $Z$ over the 50\% look-back time to $z \sim 0.8$.  
In that paper, two (of fifteen) galaxies which could
have had lower metalicities, $Z \sim 0.3 Z_{\sun}$, were tentatively identified as AGN on the
basis of their [Ne III] 3869 emission.  Our [NII] 6584 observations of these two have since
shown
however that they do in fact have the lower metallicities (see Carollo, Lilly \& Stockton 2002).

Our new larger sample reveals
a significant number with apparent $Z \le 0.5 Z_{\sun}$. 
Fig 4 shows the $M_{B,AB}$ vs. $Z/Z_{\sun}$ diagram for
the sample. The fraction of lower metallicity galaxies with $Z \le 0.5 Z_{\sun}$ depends quite
sensitively on the reddening assumed, especially in the lower $Z$ systems. With our assumed $E(B-V) \sim 0.3$
(Fig 4) it is 23\%, but it would
be reduced to only 5\% if there is no reddening at all in the lowest $Z$ objects.

\begin{figure}
\plotone{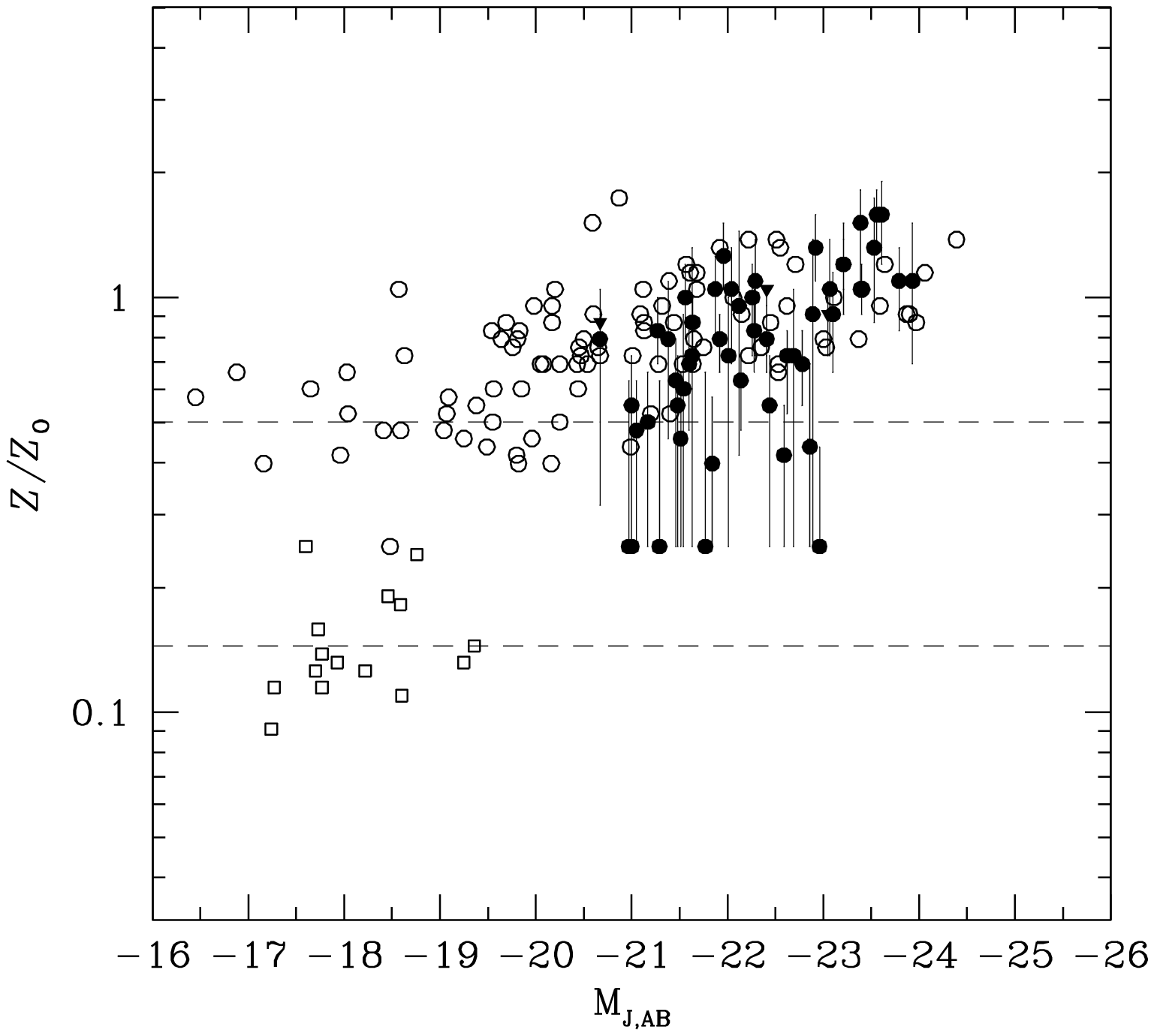}
\caption{Derived oxygen metallicities of CFRS galaxies as a function of their rest-J absolute
magnitude, compared with the JFFC sample (open symbols).  Galaxies between the dashed 
lines have uncertain 
metallicities on account of the ``turn-around'' in $R_{23}$. The clustering of objects with
one-sided errorbars at $Z = 0.25 Z_{sun}$ is artifical, stemming from the assumption that 
all objects are on the upper branch of the $R_{23}-Z$ degeneracy.  Most of the galaxies
exhibit metallicities that are very similar to the local sample at the same luminosity. 
However, a significant minority have lower metallicities, $Z \le 0.5 Z_{sun}$, which are
found in the
local sample at much fainter luminsoities. For reasons discussed in the text (see also Fig 5), 
it is unlikely that these galaxies are the progenitors 
of low metallicity galaxies today - i.e. they will migrate upwards in the diagram
to the present day rather than to the left.}
\end{figure}

Interestingly, the derived metallicity does not appear to correlate well with either
the optical morphology or with the half-light radius (Fig 5), at least for the 40\% of the
sample where we have HST images in hand. 
However, the derived metallicity does correlate with the rest-frame broad-band 
colours, especially ($B-J$) (Fig 5).  In 
addition to quite a small photospheric effect that should be associated with the lower metallicities
if shared by the young stellar population,
the trend could well be explained by
lower extinction in the lower $Z$ objects (E(B-V) $\sim 0.15$, 
modestly younger ages (e.g. 2 Gyr vs. 8 Gyr), or by
addition of a small but much younger stellar population, or some combination of all three.  
None of
these explanations would produce a large change in the stellar $M/L$ ratio.

\begin{figure}
\plotone{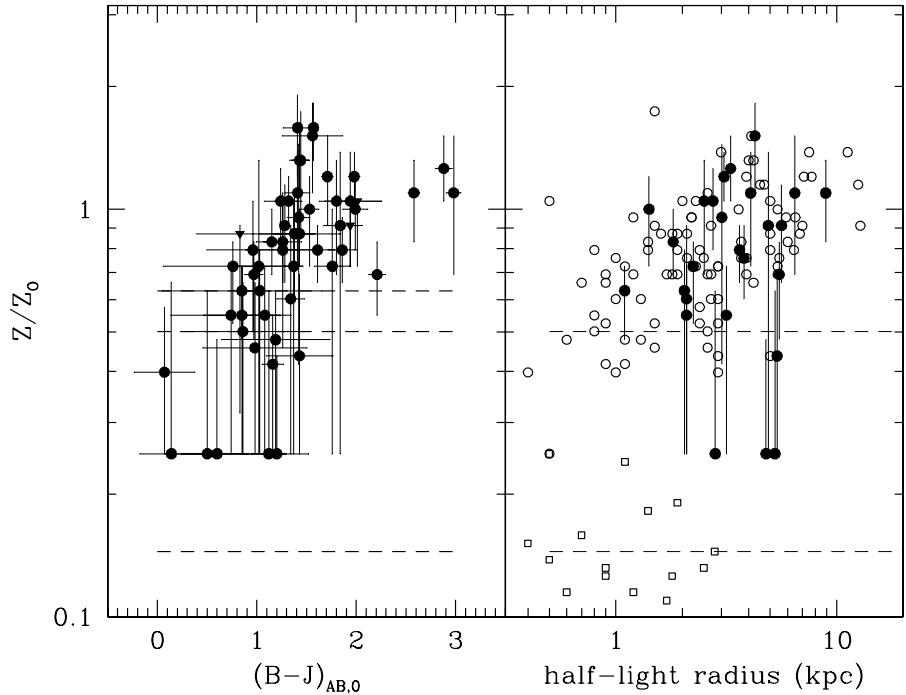}
\caption{(Left) The derived oxygen metallicity correlates quite well with overall rest-frame ($B-J$) colour. 
This is probably due to some combination of photospheric effects, lower reddening
and/or younger ages in the lower $Z$ systems, none of which would produce a large 
change in the $J$-band mass-to-light ratio.  (Right) The derived oxygen metallicity does not 
correlate well with apparent size (or morphology, not shown).  While the majority of galaxies fall 
in the area occupied by JFFC galaxies (open symbols), the low-$Z$ CFRS galaxies have larger sizes than
galaxies with corresponding metallicitiess locally.  For both reasons - normal M/L and relatively large sizes -
it is unlikely that the low-$Z$ CFRS galaxies are the progenitors of today's low-$Z$ dwarf galaxies.}
\end{figure}

The metallicity data give no support to the idea that a substantial fraction of the CFRS galaxies
(i.e. blue $L*$ galaxies at $z \sim 0.8$ - comprising the ``blue excess'' identified in the 1980's) 
will dramatically fade by several magnitudes
into today's dwarf galaxies: Most (75\% or more) are already too high metallicity ($Z/Z_{\sun} \ge 0.5$).
Those objects with dwarf-like metallicities probably, on the basis of their colours (see
Lilly et al. 2002), have a $M/L$ that is within a factor of a few of the high $Z$ systems 
and are thus unlikely to be able to
fade enough to become a present day dwarf galaxy with M$_{J,AB} > -19$.  
Where has the original ``excess'' gone? As has always been
known (e.g. discussion in Lilly, Cowie \& Gardner 1991), changing to an ${\Omega}_0 = 0.3$ 
$\Lambda$-dominated cosmology from an $\Omega = 1$ cosmology
reduces the comoving space densities at $z \sim 0.8$ 
by a factor of 2.5, and also
makes galaxies 20\% bigger and 0.5 mag more luminous.  These alone significantly reduce the
arguments for the dwarf-dominated view of galaxies at high redshifts. 
The diversity of morphologies exhibited by the galaxies in the sample suggest that we
are seeing a range of phenomena occuring in massive galaxies, including mergers,
and central star-bursts 
Bulge-forming internal disk evolution processes may also be at play
(see e.g Carollo 1999, Carollo et al 2001).

We estimate that the average H$\beta$ weighted metallicity of our sample compared 
with that of the JFFC sample (selected
with similar luminosity limits) is $\Delta$log$ Z = -0.08 \pm 0.06$.
Our uncertainty is 
currently dominated by the
uncertainty in the reddening corrections in the CFRS sample, a situation which should be rectified soon
with forthcoming $J$-band spectroscopic observations.
Extending the JFFC sample         
to the bottom of the luminosity function (weighting via the Gallego et al 1995 H$\alpha$
luminosity function) reduces the mean metallicity by only 0.08 dex, or 20\%.
Both samples at $M_{J,AB} \le -21$ 
represent a comparable fraction of the not-heavily-obscured star-formation at their
respective epochs, so it is reasonable to
suppose that our estimate for the global metallicity change could apply to star-forming
gas in the Universe as a whole.

\subsection{$R_{23}$ at $z \ge 3$ and the evolutionary history}

$R_{23}$ measurements are now available for a handful of Lyman break galaxies (LBGs) (Pettini et al. 2001).
In contrast to the CFRS galaxies that are seen at half the look-back time, all of the five LBG so far observed 
have low metallicities $Z/Z_{\sun} \le 0.5$.   Although the LBG are ultraviolet-selected, whereas both
the CFRS and JFFC samples are rest-$B$ selected, 
it is likely
that the LBG as a class dominate the galaxy population at that epoch and that a comparison between the
samples is therefore valid.

An intriguing question is whether the luminous low-$Z$ galaxies that have appeared 
in the CFRS (at the $\le 25\%$
level) by the time we get back to 
$z \sim 0.8$ are analogues of the slightly more luminous LBG that dominate the 
optical/uv-selected galaxy population at $z \sim 3$.  Inspection of Fig 3 suggests caution:
the LBG occupy a different part of the diagram than all but one of the CFRS galaxies
(the exception is CFRS22.0919)
and all of the JFFC local galaxies. The LBG exhibit much higher ionizations in their
$S_{32}$ ratio 

\begin{figure}
\plotone{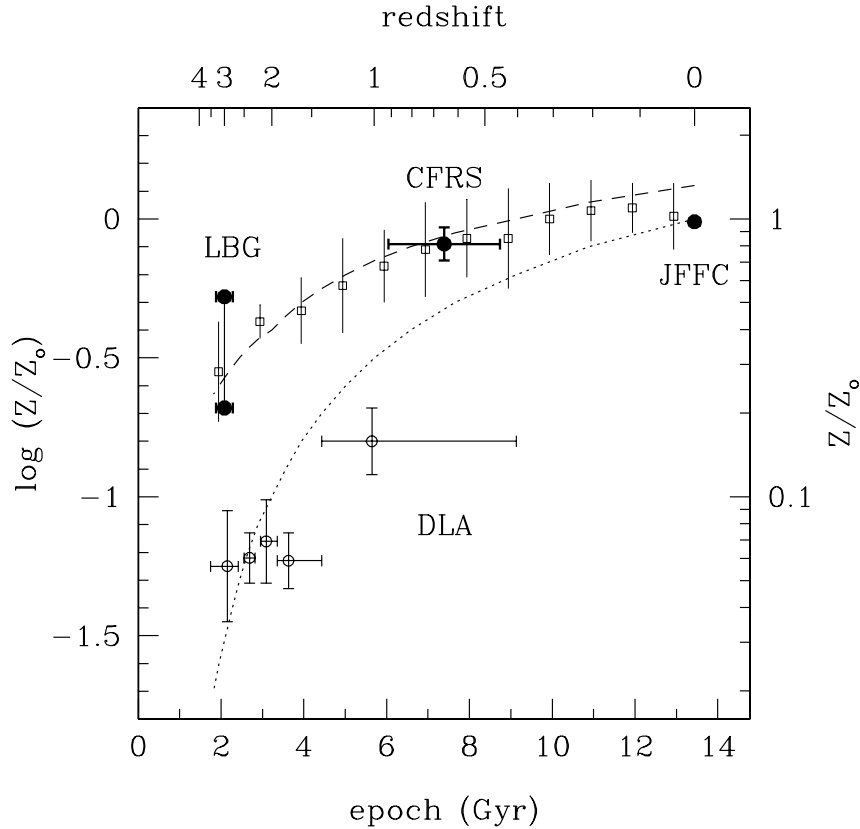}
\caption{Some evolutions of metallicity over cosmic time.  The solid symbols represent $R_{23}$-derived
oxygen abundances for star-forming galaxies - the local JFFC sample with M$_{J,AB} < -21$, the
CFRS sample of 65 galaxies described herein (both samples
H$\beta$-weighted), and the five LBG's at high redshift from Pettini et al
(2001), the last plotted at both the ``high'' and ``low'' solutions
for $R_{23}$.  Also shown as
open circles are the mean metallicities of DLA from the analysis
of Kulkarni \& Fall (2002). The evenly space open squares represent the [Fe,H] age-metallicity
relation for the solar neighbourhood from Twarog (1980).
The two theoretical models are for the global metallicity
of the ISM from Pei et al (1999) (dotted line)
and Somerville et al (2001) (dashed line).}
\end{figure}

In Fig 6, we show the average $R_{23}$-derived metallicities of the handful of LBG at $z \sim 3$, the
large sample of CFRS galaxies at $z \sim 0.8$ and the local JFFC sample. 
This is compared with Twarog's (1980) age-metallicity
relation for the Galactic disk (see the discussion in 
Garnett \& Kobulnicky 2000), Kulkarni \& Fall's (2002) analysis of
DLA metallicities, and two theoretical models:
Pei et al's (1999) ``global'' model for the metallicity of 
the ISM and Somerville, Primack \& Faber's (2001) ``collisional star-burst'' 
semi-analytic model of the galaxy population.

The agreement with Twarog's (1980) Galactic age-metallicity relation is rather striking.
The two theoretical 
models predict changes in the average metallicity back to $z \sim 0.75$ 
of $\Delta$ log$ Z \sim -0.35$ and $-0.2$
respectively.  These are larger changes than appear to be indicated by the data, and
although the discrepancy is not very significant at this point, it will probably bear
future investigation.  Fig 6 emphasizes the difference between the DLA metallicities
and those of star-forming regions.  Our lack of knowledge of the metallicities of
low redshift DLA systems means that there could be a constant offset between
them. For the time being the observational difference between galaxies and DLA
emphasizes the difficulty of modelling
the chemical evolution of individual objects and of the global Universe in a 
self-consistent way.

\section{Future directions}

Some future directions for this work are clear:

First, despite the progress, a lot of space on Fig 6 needs to be filled in, especially 
between
$1 < z < 2.5$ and, of course, at $z > 4$.  The former redshift range
is probably
when a lot of the constituents of the present-day Universe were put in place, and
understanding the development of metallicity through this epoch will be interesting.  The
range of $Z$ seen at $z \sim 0.8$ may already indicate that we are approaching an epoch
where variations in evolutionary paths are becoming important.  
Measurement of $R_{23}$ and related diagnostics will be challenging observationally
but will be quite feasible with cryogenic multi-object spectrographs on 8m 
class telescopes, at least for the lower redshifts, and with the NIRSPEC on NGST
at higher redshifts.

An extension of models of systems for which rather detailed information is available
at the present epoch (i.e. the Milky Way) so as to predict observationally-accessible
quantities such as the
average star-formation-weighted metallicity and the average neutral column density-weighted
metallicity as functions of epoch, would also help in interpreting Fig 6.  Conversely,
spatially resolved spectroscopy with integral-field spectrographs on the ground, or
in space, should be able to detect abundance gradients within systems.  Together these
would enable us to translate the impressive successes of the ``global'' analyses
to help in understanding the evolution of individual objects.

Finally, our understanding the physical meaning of integrated spectra of objects (e.g. the
ionization differences between the LBG at $z \sim 3$ and superficially similar objects at
lower redshifts) needs to be developed.

\acknowledgments We thank Jarle Brinchmann, Tracy Webb and Mark Brodwin for their
assistance in compiling the infrared photometry for the CFRS part of this program, 
some of which was from
unpublished data.

\end{document}